\documentclass[twocolumn,pra,showpacs,amsmath,amssymb,superscriptaddress]{revtex4}

\usepackage{amsmath,amsfonts,amsthm,graphicx,color}

\newcommand{\ket}[1]{|#1\rangle}
\newcommand{\bra}[1]{\langle #1|}
\newcommand{\beq} {\begin{equation}}
\newcommand{\enq} {\end{equation}}
\newcommand{\ber} {\begin {eqnarray}}
\newcommand{\enr} {\end {eqnarray}}

\newcommand {\er}[1] {equation (\ref{#1}) }

\newcommand{\diag}{\mathop{\mathrm{diag}}}
\newcommand{\CC}{\mathbb{C}}

\def\one{\leavevmode\hbox{\small1\normalsize\kern-.33em1}}

\begin{document}

\title{Discrimination between evolution operators}
\author{T. V\'ertesi}
\email{tvertesi@dtp.atomki.hu}
\affiliation{Institute of Nuclear Research of the Hungarian Academy of Sciences\\
H-4001 Debrecen, P.O. Box 51, Hungary}
\author{R. Englman}
\email{englman@vms.huji.ac.il}
\affiliation{Soreq NRC,~Yavne~81800,~Israel}

\begin{abstract}
Under broad conditions, evolutions due to {\it two}  different
Hamiltonians are shown to lead at some moment to  orthogonal
states. For two spin-$\frac{1}{2}$ systems subject to precession
by different magnetic fields the achievement of orthogonalization
is demonstrated for every scenario but a special one. This
discrimination between evolutions is experimentally much simpler
than procedures proposed earlier based on either sequential or parallel application
of the unknown unitaries. A lower bound for the orthogonalization time is
proposed in terms of the properties of the two Hamiltonians.

\end{abstract}


\maketitle


\section {Introduction}

Quantum information is predicated on the preparation and
identification of evolving quantum states; consequently, the
determination of quantum states is of prime importance (\cite
{Diosi}, Chapter 6). Perfect determination is possible only for
states that are known to be orthogonal \cite{BennettW, Chefles};
in the absence of such knowledge the efficiency of imperfect
determination can be improved by making non-projective
and/or collective measurements whose outcomes are associated with entangled states,
or by exploiting the quantum mechanical interference between states \cite {Meyer}.
Still, for perfect determination of non-orthogonal (pure or mixed)
states an infinite number of state-copies are needed. Recently a
probabilistic measure of the discriminatory process was proposed
\cite {AudenaertCMBMAV}.

 The above picture changes when  discrimination is  made
between unitary operations; for these a finite number of runs of
the unknown gates suffice \cite{Acin}-\cite{DuanFY}, a fact which
is by itself surprising, since the {\it number} of independent
parameters entering the unitary operations is of the order of the
square of the number of states (assuming the latter to be finite).
What balances this circumstance, is that there is a choice of
states on which the unitaries can be tested. It now seems safe to
state that while in certain cases the entanglement property of
this chosen state can aid in the discrimination, perfect
discrimination is also possible without entanglement
\cite{DuanFY}.

The present work addresses unitaries in the framework  of the
class of evolution operators and, in particular, those with
time-independent Hamiltonians $H$, so that their form is
$e^{-iHt}$. Here the time $t$ takes over in many senses the role
of the (discrete) number $N$ of the unitary gates featured in the
above mentioned works \cite{Acin}-\cite{DuanFY}, but in terms of
practical effort the automatically time-evolving unitary is much
preferable to repeated subjection to gates. The aim of this work
is thus the determination of a minimal discrimination time between
two different evolution operators, with  liberty in the choice of
the starting state. We identify the discrimination time with the
time (to be denoted by $t_{\perp}$) that  two differently evolving
states first become orthogonal. An algebraic treatment of the
opposite issue, the recurrence of states under pulsed and other
time dependent perturbations, was given in \cite{Mielnik} and
\cite {FernandezM}.

The relationship of this topic to time-orthogonality of states and
to the rate of sweeping the Hilbert space is evident, the latter
issue having  already been solved for some states in \cite
{MargolusL}-\cite{CarliniHKO} and endowed with geometrical
interpretation in \cite{BrodyH}-\cite{BenderBJM}. In these, lower
bound expressions are given in terms of the average energy
$\overline{E}$ and the mean deviation of the energy $\Delta E$ of
the state. The states under present consideration are
multi-component pure states, for which the existence of
orthogonality times is studied in section~III. The obvious
difference (noted in section~IV) with the previous references  is
that we have two different (in general, non-commuting)
Hamiltonians, rather than a single one. A single-qubit state
example is fully worked out in section~V. The lower bounds
$t_{lb}$  for $t_{\perp}$ derived in section~6 by two different
methods are supported by the solved example (of section~V). The
fastest orthogonalizing pair of Hamiltonians turns out to be a
rather trivial couple (section~VI~C). A discussion in section~VII is
mainly devoted to entanglement.

\section{Two Hamiltonians}

 We consider two Hamiltonians $H_a$ and $H_b$ that give rise to  the
 evolution operators $e^{-A(t)}$ and $e^{-B(t)}$ . Here $A(t)=iH_a t$
 and $B(t)=iH_b t$, where $t$ is time. Physically, $H_a$ can be made
 different from $H_b$ by adding different excitation operators
 to the common system Hamiltonian (say, $H_0$).
 One can discriminate between the two Hamiltonians (or excitation mechanisms)
 if, for some initial state $\psi$, $e^{-B(t)}\psi$ is
 orthogonal to $e^{-A(t)}\psi$ at some orthogonality time (or
 times) $t=t_{\perp}~ $\cite{DarianoPP}-\cite {DuanFY}. This means
 \beq{\bra\psi}e^{B(t_{\perp})}e^{-A(t_{\perp})}{\ket\psi}={\bra\psi}e^{iH_b
 t_{\perp}}e^{-iH_at_{\perp})}{\ket\psi}=0.\label{braket}\enq

\section{Derivation of Orthogonality}
Since $e^{-A(T)}$ and $e^{B(T)}$ are both unitary, their product
\beq e^{C(t)}=e^{B(t)}e^{-A(t)}\label{C}\enq is also unitary and
can therefore be brought to the form
$\diag(e^{i\theta_1(t)},e^{i\theta_2(t)},\ldots,e^{i\theta_d(t)})$, by
a unitary transformation matrix $M(t)$  in which  all the
$\theta$'s are real ($d$ is the dimension of the system's
Hilbert space). When $A(t)$ and $B(t)$ are different (i.e. $A(t)$
is not of the form: $B(t)+$ a Hilbert-space scalar, which is possibly time
dependent), not all the $\theta$'s are the same and $M(t)$ differs
from the unit matrix. (In the converse case, when $A(t)$ and
$B(t)$ are not different, all $\theta(t)$'s are the same and
$M(t)$ is the unit matrix.) Suppose now that two $\theta$'s that
differ have indices $i$ and $j$.

We now construct the state \beq
\psi(t)=M^{+}(t)v(i,0;j,\alpha)/\sqrt{2}\label{os}\enq in which
$v(i,0;j, \alpha)$ is a $d$-dimensional column vector whose $i$'th
element is $e^{i0}=1$ and whose $j$'th element is $e^{i\alpha}$
($\alpha$ being real), all other entries in the column being zero.

We now form the bracket \ber {\bra\psi}e^{B(t)}e^{-A(t)}{\ket\psi}
& =
 & {\bra\psi}M^{+}(t) \diag(e^{i\theta_1(t)},e^{i\theta_2(t)},\ldots,
e^{i\theta_d(t)}) \nonumber\\
&& \times M(t) {\ket\psi}\label{brak1}\\
& = & \frac{1}{2} ( e^{i\theta_i(t)}+ e^{i\theta_j(t)})\label{brak2}\enr
Since the $\theta$'s differ (= the cases when $A(t)$ and $B(t)$
are different), it is expected that for some $t=t_{\perp}$ they
will differ by $\pi$ and the bracket will vanish.

The reason that the state $\psi$ has only two non-zero components
can be explained as follows. Since \beq
{\bra\psi}\diag(e^{i\theta_j(t)}){\ket\psi}= \sum_j
e^{i\theta_j(t)}|\psi_j|^2\label{diag2}\enq and the numbers
$|\psi_j|^2$ form a discrete probability distribution, the set of
possible values of ${\bra\psi}\diag(e^{i\theta_j(t)}){\ket\psi}$
forms a convex set in the complex plane, with the numbers
$e^{i\theta_j}$ being extremal points. Initially, this set is the
singleton $\{1\}$. With increasing time, the size of the set
increases. Orthogonality is first achieved when the set includes
the point $0$. This happens when $0$ hits one of the edges of the
convex set. The edges correspond to $\psi$ with at most two
non-zero components.

\section{Evolution-operator Analogy}
Writing the (unitary) exponential $e^{C(t)}$ introduced in \er{C}
as $e^{-it[\frac{-C(t)}{it}]}$, the quantity in the square bracket
is not time independent. This is seen by recalling the
Campbell-Baker-Hausdorff formula for $C(t)\equiv
\ln[e^{B(t)}e^{-A(t)}]$) that contains commutators of $A(t)$ and
$B(t)$ to arbitrary order. However, to the zeroth order of
commutators, or equivalently for the linear-in-$t$ approximation
for $C(t)$, this quantity is a "Hamiltonian", it being the
difference of two Hamiltonians. To this order, therefore,
$e^{C(t)}$ is an evolution operator. The speed at which a
Hamiltonian evolution operator develops a state into one or
several orthogonal states has been studied previously (e.g., \cite
{MargolusL} and references therein). One result is that the time
$t_{\perp}$ at which an initial state (starting at zero energy)
evolves to an orthogonal state (in units for which $\hbar=1$)
is given by \beq t_{\perp}\geq \frac{\pi}{{2\bar E}},\label{lb}\enq
where ${\bar E}$ is the average energy.

The use of this (approximate) measure can be employed in various
situations, for bipartite and multipartite systems (entangled and
otherwise). We shall also use it in the following illustrative
example.
\section{ Illustration: a qubit}

Let us write out the Hamiltonians for the evolutions  $a$ and $b$
explicitly,
\begin{eqnarray}
H_a &=& r_{0a}I+\omega_a\sum_{k=x,y,z}r_{ia}\sigma_k, \nonumber \\
H_b &=& r_{0b}I+\omega_b\sum_{k=x,y,z}r_{ib}\sigma_k. \label{Hab}
\end{eqnarray}
Then according to Sec.~II, we can write $A(t)=iH_a t$ and
$B(t)=iH_b t$, where $t$ denotes time. The formula which we intend
to verify is Eq.~(1), namely, for any given pair of $H_a$ and
$H_b$ we choose $t=t_{\perp}$ appropriately so that Eq.~(1) is
fulfilled. Next we omit the $r_0$ part from Eq.~(\ref{Hab}) since
this induces merely a global phase shift in the evolution of the
state, which has no effect on the left hand side of Eq.~(1).

The problem is an analog of a spin-$1/2$ particle precessing with
frequency $\omega_a$ for a time $t=t_{\perp}$ in a magnetic field
pointing in the direction $\vec r_a=[r_{xa},r_{ya},r_{za}]$ (where
the corresponding Larmor frequency $\omega_a$ is proportional to
the magnetic field strength) and subsequently precessing about
$\vec r_b=-[r_{xb},r_{yb},r_{zb}]$ for the same time interval
$t=t_{\perp}$ with a frequency $\omega_b$. $|\vec r_a|=|\vec
r_b|=1$. The question is, whether the final state $\psi_f\equiv\psi(t=2t_{\perp})$
becomes orthogonal to the initial state $\psi\equiv\psi(t=0)$ for a suitable $t_{\perp}$.
Visualizing this situation within the Poincare-sphere description, we ask whether
the resulting state $\psi_f$ points to an antipodal point on the
sphere, meaning $\langle \psi_f|\psi\rangle=0$. Thus the problem
can be converted to a problem of purely geometrical origin.

Nevertheless, let us remain at the quantum-mechanical level, and
 express the rotation matrices $\exp(-iH_a t)$ and $\exp(iH_b t)$
 by Pauli matrices, by applying the following identity:
\begin{eqnarray}
R_{\vec r}(\theta) &\equiv& \exp(-i\theta \vec r \vec \sigma/2) =
\cos(\theta/2)\one \nonumber\\ && -i\sin(\theta/2)(r_x\sigma_x +
r_y\sigma_y+r_z\sigma_z), \label{Rab}
\end{eqnarray}
where the operator $R_{\vec r}(\theta)$ in the Poincare picture defines
a rotation by $\theta$ about the axis $\vec r$.

After some algebra we can express the product of the two operators
$R_{\vec r_b}(\theta_b)$ and $R_{\vec r_a}(-\theta_a)$ by one
operator $R_{\vec r_{ab}}(\theta_{ab}) = R_{\vec r_b}(\theta_b)
R_{\vec r_a}(-\theta_a)$ \cite{NielsenC}, where the overall angle
$\theta_{ab}$ and the new axis $\vec r_{ab}$ are given by
\begin{eqnarray}
\cos\left(\frac{\theta_{ab}}{2}\right) &=& \cos(-\theta_a/2)\cos(\theta_b/2)\nonumber \\&&-\sin(-\theta_a/2)\sin(\theta_b/2)\vec r_a\cdot \vec r_b, \label{cos} \\
\sin\left(\frac{\theta_{ab}}{2}\right)\vec r_{ab} &=& \sin(-\theta_a/2)\cos(\theta_b/2)\vec r_a\nonumber \\ &&+ \cos(-\theta_a/2)\sin(\theta_b/2)\vec r_b \nonumber\\
&&-\sin(-\theta_a/2)\sin(\theta_b/2)\vec r_b \times \vec r_a.
\label{newrot}
\end{eqnarray}
Let us relabel the Poincare-sphere so that $\vec r_{ab}$ points to
the north pole.
 In this case, the imaginary part of \er{Rab} is simply $-i\sin(\frac{\theta_{ab}}{2})|r_{ab}|\sigma_z$ and
 in order to satisfy the imaginary part of $\langle \psi_f|\psi\rangle=0$,
  the initial state  must take the form \beq \psi\propto[|\uparrow\rangle + e^{i\alpha}|\downarrow\rangle\label{psi2}]\enq
  Then $\langle \psi|\sigma_z|\psi\rangle=0$.

  We now turn to the second task:
   to nullify  the contribution from the real part of \er{Rab}. Let us call $\gamma$ the angle between
   $\vec r_a$ and $\vec r_b$  . Then we have in \er{cos}
   $\vec r_a \cdot \vec r_b = \cos\gamma$.
      With \er{cos} in hand we may write the real part of the orthogonality condition
    in the form
\begin{equation}
0 =
\cos(-\theta_a/2)\cos(\theta_b/2)-\sin(-\theta_a/2)\sin(\theta_b/2)\cos(\gamma).\label{crit1}
\end{equation}
With further algebra and observing that in our case
$\theta_a=2\omega_a t$ and $\theta_b=2\omega_b t$, we arrive at
the form
\begin{equation}
0=\cos^2\frac{\gamma}{2}\cos(\omega_a-\omega_b)t+\sin^2\frac{\gamma}{2}\cos(\omega_a+\omega_b)t.
\label{crit}
\end{equation}
Let us pick from the right hand side (RHS) of this equation the
term with greater amplitude. At $t=0$ the RHS is equal to $1$.
Then if $\cos^2(\gamma/2)>\sin^2(\gamma/2)$, at
$t=t_{\perp}^{max}=|\pi/(\omega_a-\omega_b)|$ or else (by the
converse inequality) at
$t=t_{\perp}^{max}=\pi/(\omega_a+\omega_b)$ the expression on the
right must be negative. Thus by continuity argument there must be
a point ($t_{\perp}$) between $t=0$ and $t=t_{\perp}^{max}$ where
the RHS of Eq.~(\ref{crit}) becomes zero. However if
$\cos^2(\gamma/2)>\sin^2(\gamma/2)$ that is $\gamma<\pi/2$ and
$\omega_a=\omega_b$, $t_{\perp}^{max}$ becomes infinite, and the
condition $\langle \psi_f|\psi\rangle=0$ is not guaranteed to be
fulfilled.

This result has the following implications. If $\omega_a \neq
\omega_b$, then the Hamiltonians $a$ and $b$ could be perfectly
discriminated. However, if $\omega_a=\omega_b$ and the angle
$\gamma$ between $\vec r_a$ and $\vec r_b$ is smaller than $\pi/2$
then Eq.~(1) cannot be fulfilled for any finite time $t_{\perp}$.
When $\omega_a \neq \omega_b$ but $\vec r_a=\vec r_b$, the two
Hamiltonian operators are different (by the definition given) and
they can still be perfectly distinguished within a finite time
$t$, although  they commute. In \cite{DarianoPP, DuanFY} it is the
run number $N$, which is broadly equivalent to the orthogonality
time in the present work, that tends to infinity when the two
unitaries become the same. [A formal expansion of \er{crit} for
short times gives \beq
t^2=\frac{8}{\omega^2_a+\omega^2_b-2\omega_a \omega_b \vec
r_a\cdot \vec r_b}\label{shortt}\enq and shows this tendency (but
the short time expansion is no longer valid)]
 \subsection{Evolution time approach}
 The difference Hamiltonian, which as noted previously, can be
 viewed as approximately guiding the initial state to an
 orthogonal state, is \beq H_a-H_b= (\omega_a \vec r_a-\omega_b \vec
 r_b)\cdot{\vec \sigma}.\label{diffH}\enq This can be diagonalized
 to give the average energy ${\bar E}$ of the two eigen-states
 (with the lower state being placed at zero energy) given by \beq
{\bar E}=\sqrt{\omega^2_a+\omega^2_b - 2\omega_a\omega_b \vec
r_a\cdot \vec r_b }.\label{En}\enq

\begin{figure}
\includegraphics[width=8cm]{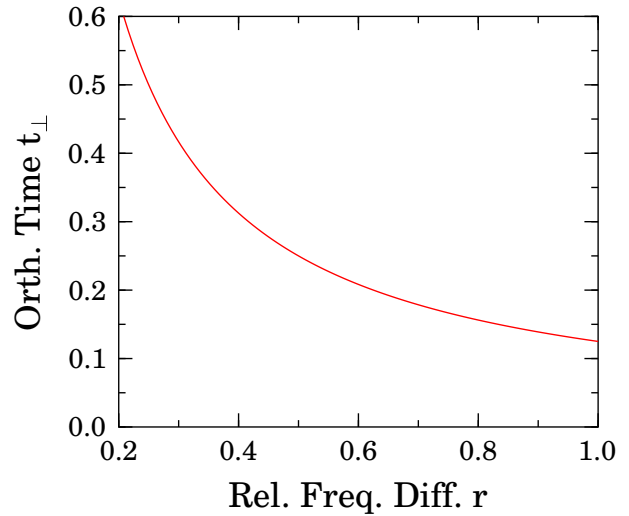}\\
\caption{Orthogonality time in a qubit against relative frequency
difference $r$ (defined in \er{r1}) of the precessing magnetic
fields, for perfect  alignment ($\gamma=0$). The curve obtained
from the solution of \er{crit1} is indistinguishable from the
evolutionary orthogonality time in the RHS of \er{lb}.  The time
is in inverse units of the average precession radian frequencies
$\frac{\omega_a+\omega_b}{4\pi}$.}
 \label{orthtime0}\end{figure}

We first show results for a case when the evolution time approach
is expected to give orthogonality times identical to the unitary
operator orthogonality times. This will be the case when the two
magnetic fields are perfectly aligned: \beq \vec r_a\cdot \vec
r_b\equiv \cos\gamma =1~,~~ \gamma=0.\label{aligned}\enq The
fields' Larmor frequencies still differ,\beq \frac{\omega_a
-\omega_b}{\omega_a +\omega_b}\equiv r \neq 0\label{r1}\enq so
that while the $a$ and $b$ Hamiltonians commute, the evolution
operators are still different. The computed orthogonality times
($t_{\perp}$), obtained from solution of \er{crit}, as function of
the normalized frequency differences $r$  are shown in Figure~1.
When we compare them to the approximate (lower bound) expression
given in the RHS of the inequality~(5), using the above formula
for ${\bar E}$, we obtain perfect agreement (without adjustment of
a proportionality factor).

In the next figure (Fig.~2) we show (with the dotted curve) the
computed orthogonality times, again obtained from solution of
\er{crit}, as a function of the alignment angle $\gamma$, with
the frequency difference fixed (at $r=0.5$) and compare this to the
approximate (lower bound) expression shown in \er{lb}(the full
curve in Fig.~2). There is still a reasonable agreement between
the two curves. However, at non-perfect alignment angles
($|\gamma|>0$) this agreement gets spoiled for frequency
differences smaller than $r=0.5$ or $\frac{\omega_a}{\omega_b}<3$.
\begin{figure}
\includegraphics[width=8cm]{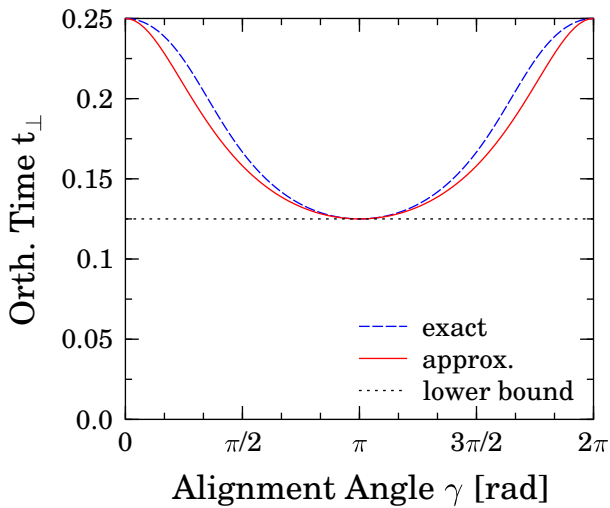}
\caption{Orthogonality time in a qubit against alignment angle
$\gamma$ between the precessing magnetic fields. ~Dashed curve:
solution of \er{crit1}.~~Full curve: from approximate evolutionary
orthogonality formula \er{lb}.~Dotted line: lower bound obtained
from \er{bound2}. The time is in inverse units of  the average
radian frequency $\frac{\omega_a+\omega_b}{4\pi}$. The ratio of
the frequencies was fixed at $\frac{\omega_a}{\omega_b}=3$. For
negative angles the curves are the reflections of those shown.}
\label{orthtime1}\end{figure}
\section{Lower Bound on the Orthogonality Time}
\subsection{Derivation from the Anandan-Aharonov  relation \cite{AnandanA}}
We derive a lower bound on the orthogonality time $t=t_{\perp}$
for general $d$-dimensional systems, introduced in Section~II. For
that purpose we invoke the geometric quantity \cite{AnandanA}
\begin{equation}
s=2\int{\Delta E(t)dt}, \label{s}
\end{equation}
where $\Delta E(t)$ is the energy uncertainty in state $\psi$
defined by
\begin{equation}
\Delta E= \sqrt {\langle\psi|H^2|\psi\rangle - \langle
\psi|H|\psi\rangle ^2}.
\end{equation}
According to \cite{AnandanA}, the quantity $s$ is the distance
along a given curve $C$ in the projective Hilbert space $P$ as
measured by the Fubini-Study metric. In Section~II, \er{braket} can
be interpreted as state $\psi$ evolving for a time interval
$t_\perp$ under Hamiltonian $H_a$ and, following that, for the
same interval $t_\perp$ under Hamiltonian $-H_b$ to an orthogonal
state $\psi_f$. Thus one is allowed to write the right hand side
of equation~(\ref{s}) as two separate time integrals, whose sum is
$2\Delta E_a t_\perp + 2\Delta E_b t_\perp$ in our particular
case. On the other hand, the shortest distance between orthogonal
states can be achieved along a geodesic curve where $s=\pi$, hence
in a generic evolution we have the relation $s\geq \pi$. Summing
up these facts and substituting them into equation~(\ref{s}) we
obtain the lower bound
\begin{equation}
t_\perp \geq \frac{\pi}{2(\Delta E_a +\Delta E_b)} \equiv t_{lb}
\label{bound1}
\end{equation}
on the orthogonality time needed to discriminate perfectly between
two evolution operators $e^{-A(t)}$ and $e^{-B(t)}$ defined in
Section~II.

Now let us suppose that we have knowledge only about the
eigenvalues of the Hamiltonians $H_a$ and $H_b$ in question.
Having the eigenvalues of a $d$-dimensional Hamiltonian $H'$
(where $d$ can not be infinite but can be arbitrarily large) gives
rise to the upper bound $2\Delta E'\leq E^{'max}- E^{'min}\equiv
2\omega'$, where $E^{'max}$ and $E^{'min}$ denote, respectively, the
highest and lowest eigenvalues of the Hamiltonian $H'$. The
preceding upper bound on insertion into equation~(\ref{bound1})
yields the inequality
\begin{equation}
t_\perp \geq \frac{\pi}{2\omega_a + 2\omega_b} \label{bound2}
\end{equation}
for the time $t_\perp$ to discriminate between the pair of
evolution operators $e^{-A(t)}$ and $e^{-B(t)}$. Note, that this
bound on $t_\perp$ is sharp in the sense that, given $\omega_a$
and $\omega_b$, we can always construct appropriate Hamiltonians
$H_a$ and $H_b$ from them so that the bound~(\ref{bound2}) for a
given initial state $\psi$ would be saturated. This can be
achieved by choosing, in particular, the Hamiltonians
\begin{eqnarray}
H_a &=& \omega_a|E_+\rangle\langle E_+| - \omega_a|E_-\rangle\langle E_-| \;, \nonumber \\
H_b &=& -\omega_b|E_+\rangle\langle E_+| + \omega_b|E_-\rangle\langle E_-| \;, \label{Hamab}
\end{eqnarray}
and letting the initial state $|\psi\rangle = \frac{1}{\sqrt
2}(|E_+\rangle + e^{i\alpha}|E_-\rangle)$ for some $\alpha \in
[0,2\pi]$. We obtain for the special studied case $d=2$ in
Section~V, that the optimal $t_\perp$ is obtained for the
alignment angle $\gamma=\pi$, a choice that indeed generates a pair
of Hamiltonians of the kind corresponding to
equation~(\ref{Hamab}).

\subsection{Derivation through Brody et al's formula
\cite{BrodyH,BenderBJM}} Actually, in order to obtain the
bound~(\ref{bound2}) we do not need to resort to the
Anandan-Aharonov relation~(\ref{s}), as we will show in the
following by using formulas from Refs.~\cite{BrodyH} and
\cite{BenderBJM}, whose derivation only requires elementary
trigonometry. The problem discussed in these works is the
following: Consider a pair of initial and final states $\psi_I$
and $\psi_F$ in a $d$-dimensional Hilbert space. The task is to
find the Hamiltonian $H$ on this Hilbert space which takes
$\psi_I$ into $\psi_F$ in the shortest possible time $\tau$. It
has been found by elementary considerations (e.g., Eq.~(5) in
Ref.~(\cite{BenderBJM})), that
\begin{equation}
\tau = \frac{2\arccos|\langle\psi_I|\psi_F\rangle|}{2\omega}\;,
\label{tau}
\end{equation}
where $2\omega = E^{max}-E^{min}$ denotes the difference of the
largest and the smallest eigenvalues of $H$. In applying the above
formula~(\ref{tau}) for our situation we need to involve an
intermediate state $\psi_m$, so that $|\psi_m\rangle = e^{-i H_a
t_\perp}|\psi\rangle$ and $|\psi_f\rangle = e^{i H_b
t_\perp}|\psi_m\rangle$. Rearranging equation~(\ref{tau}) we
obtain two equations involving our three states,
\begin{eqnarray}
 \alpha_{a} \equiv 2\arccos|\langle\psi|\psi_m\rangle| &=& t_\perp 2\omega_a\;, \nonumber\\
 \alpha_{b} \equiv 2\arccos|\langle\psi_m|\psi_f\rangle| &=& t_\perp 2\omega_b \;.
\label{tauab}
\end{eqnarray}
If we add up the two angles, $\alpha_a$ and $\alpha_b$, on the
left hand side of equation~(\ref{tauab}), since $\langle
\psi|\psi_f\rangle =0$, the sum can minimally take up the value
$\pi$. Then we have $t_\perp (2\omega_a + 2\omega_b) = \alpha_a +
\alpha_b \geq \pi $. That is, $t_\perp\geq \pi/(2\omega_a +
2\omega_b)$; thus we are back to the inequality~(\ref{bound2})
derived earlier with the aid of the Anandan-Aharonov relation.
However, from the present derivation it is more transparent that
in order to achieve the smallest value for $t_\perp$, it is necessary
to express $\psi_m$, using some linear combination of the
initial and final states $\psi$ and $\psi_f$ (otherwise the angle
$\alpha_a + \alpha_b$ would be greater than $\pi$). Since for a
generic pair of Hamiltonians $H_a$ and $H_b$ with dimensions $d>2$
the condition that $\psi_m$ is in the two-dimensional subspace
spanned by $\{|\psi\rangle,|\psi_f\rangle\}$ is a very severe
condition to meet, the orthogonality time $t_\perp$ would in
general be much larger than the value given by the lower
bound~(\ref{bound2}).

\subsection {Finding Hamiltonians for the lower bound}
This subsection generalizes the argument leading to \er{Hamab}.

We consider two Hamiltonians $H_a$ and $H_b$ that give rise to the
 evolution operators $e^{-A(t)}$ and $e^{-B(t)}$ . As before, $A(t)=iH_a t$
 and $B(t)=iH_b t$. We write $e^{B(t)}e^{-A(t)}
 \equiv e^{-C(t)}\equiv e^{-iH_c(t)t}$. We assume to know orthogonality lower
 bounds  $t_{lb}^a \equiv t_{lb}(E_a)$ for $H_a$ alone (and similarly $t_{lb}(E_b)$ for
 $H_b$) as given
 in previous work \cite{DuanFY}-\cite{MargolusL}, expressed in terms of some
 energy $E_{a,b}$(e.g., the span of the energy spectrum, or the energy
 uncertainty, as in \er{s}). Having the Hamiltonian $H_a$ we now  try to find
 another Hamiltonian $H_b$ such that will achieve the lower bound
 of the orthogonality time. [We shall also trivially find the upper bound
 ($\infty$) orthogonality time].

 We recall from the Campbell-Baker-Hausdorff expansion that one can express
$-C(t)\equiv \ln[e^{B(t)}e^{-A(t)}]$ or  $-it H_c(t)\equiv
\ln[e^{iH_b}e^{-iH_at}]$ in terms of a hierarchy of commutators,
the zero order term being $H_c(t)\approx H_a-H_b\equiv H_c$. We
also assume  that $H_a$ is given and is such that $TrH_a=0$ (which
should be possible to construct for Hamiltonians having a finite
basis) and consider only traceless $H_b$ ($TrH_b=0$). We further
set $Tr (H^2_b)=K^2 Tr (H^2_a)$, where $K$ is an arbitrary constant.
There may be some terms in $H_a$ and $H_b$ that are common to both
of them and commute with both (e.g., kinetic energy operators that
commute with spin variables, where only the latter constitute the
differing parts in $H_a$ and $H_b$.) We discount these terms.

Then we  conjecture  that $Tr (H^2_c(t))$ is maximal when $H_b=
KH_a$, where $K$ is negative. In \er{Hamab},
$K=-\frac{\omega_b}{\omega_a}$. (A proof for this conjecture,
provided to us by a referee, is given with slight modifications in
the Appendix.)

Then $H_c=(1+|K|)H_a= H_a-H_b$, $Tr (H^2_c(t))=(1+|K|)^2Tr
(H^2_a)$, and the lower bound for $c$ is
$t_{lb}^c=t_{lb}(E_a+|E_b|)\equiv t_{lb}(E_c)$.(We write the
modulus of $E_b$, since only positive energies enter in the
expressions $t_{lb}(E)$, whereas with $K$ negative $E_b$ could be
misinterpreted as a negative term.)

Also, trivially, when $K=1$, $H_c=0$ and the resulting bra-ket is
always unity, never zero. This choice (not necessarily unique)
achieves the infinite orthogonality time upper bound.
\section {Summary and Conclusions}
This paper treated the following: Two different Hamiltonians
($H_a$ and $H_b$) are given and it is desired to be able to
distinguish between them. The method of this paper is to form
unitary operators out of these Hamiltonians and apply them on an
initial state. Should the states become orthogonal, this indicates
the distinctness of the Hamiltonians. However, two issues emerge:
First , there is no guarantee that for any chosen initial state
differently evolving states ever become orthogonal. Secondly
(which has some practical implications), what are the times
(possibly minimal times) within which one can expect orthogonality to be
obtain? By constructing some {\it special} states out of $H_a$ and $H_b$,
as in section~III and shown in \er{os}, one can expect to
reach orthogonality. The construction is feasible for systems with
arbitrary Hilbert-space dimensionality (and for any number of
components), but the achievement of orthogonality is not
guaranteed in this work. However, the illustration worked out in
section~V for a qubit supports the existence of orthogonality,
except for effectively identical Hamiltonians (such as when the
non-scalar parts of $H_a$ and $H_b$ are proportional to each
other). Our formal apparatus has indicated such special cases by
infinitely long orthogonality times, as in the text after
\er{crit}.

 This paper, which has quested discrimination between two Hamiltonians,
  can be regarded as an
intermediate between past works treating state-orthogonalizations
and unitary-discriminations. We have found ways to establish
orthogonalization times $t_{\perp}$ in general pure-state
situations and calculated $t_{\perp}$ in a single qubit model.
Lower bounds $t_{lb}$ of $t_{\perp}$ were proposed in terms of
quantities arising from the two Hamiltonians.

It is not clear whether the orthogonality times can be shortened
by starting with entangled states (or whether infinite
$t_{\perp}$'s can be reduced to finite values with entangled
states). From the way that maximal distinguishability is found for
single qubit states in \cite{Acin} (following eq. (12) there),
{\it either} through maximal entanglement {\it or} by optimal
state orientation (represented there by $\vec s$), one would
expect that the $t_{\perp}$ result found for optimal state will
not be further reduced with entangled states.

\subsection*{Acknowledgements}

The authors are indebted to an anonymous referee for the
explanation at the end of section~III and for providing the proof in
the Appendix for the conjecture stated in section~VI~C.
T.V. would like to thank Dr.\hspace{-0.1cm} K\'aroly~F. P\'al
for several useful discussions. T.V. was supported by a Grant \"Oveges of the
National Office for Research and Technology.

\section*{Appendix: Proof of Conjecture}

The following theorem  immediately establishes the assertion in
section~VI~C of the text that with $iH_c(t)t =\ln(e^{iH_a
t}e^{-iH_bt})$ and the ratio $Tr(H_b^2)/Tr(H_a^2)$ fixed, $Tr
(H^2_c(t))$ is maximal when $H_b= KH_a$, where $K$ is negative.

Theorem 1.
{\it Let the complex logarithm be defined on $\CC\setminus R^-$,
i.e., with a cut along along the negative real axis.
For all unitaries U and V \beq \|\ln(UV)\|_2\leq \|\ln U\|_2+\|\ln V
\|_2,\label{UV1}\enq}where $\|~\|_2$ denotes the square root of the
trace of the matrix squared (the Frobenius norm).
An alternative statement of the theorem is, in terms of Hermitian $A$ and $B$
with spectrum in the half-open interval $(-\pi,\pi], $
\beq\|\ln(e^{iA}e^{iB})\|_2\leq \|A\|_2+\|B\| _2.\label{AB1}\enq In
the notation of the conjecture, Theorem 1 states that for $U
=e^{iH_a t}$, $V=e^{-iH_bt}$ and $H_c(t)$ as above, \beq
\|H_c(t)\|_2\equiv\|\ln(e^{iH_a t}e^{-iH_bt})\|_2\leq
\|H_at\|_2+\|H_b t \|_2.\label{AB2}\enq But when $H_b=KH_a$, with
$K=-\|H_b\|_2/\|H_a\|_2$, \beq
\|H_c(t)t\|_2=(1-K)\|H_at\|_{2}=\|H_at\|_2+\|H_bt\|_2,\label{Hc3}\enq
which satisfies the equality option in \er{AB2}, showing that
the  choice made, $H_b=KH_a$, maximizes $Tr (H^2_c(t))$. In the
proof of the theorem (which is done by  induction) the time $t$ is
irrelevant, so we replace the three Hamiltonians in the text by
three related Hermitian matrices designated as $X$,$Y$,$Z(1)$.
These are connected through \beq e^{iZ(s)}=e^{iX}e^{isY}~~~(0\leq
s\leq 1)\label{XYZ}\enq with the eigenvalues of $Z(s)$ restricted
to the half-open interval $(-\pi,\pi]$. In addition, $Z(s)$ is
constrained to be continuous over $0\leq s\leq 1$, which requires
that $e^{iX}e^{isY}$ has no eigenvalue equal to $-1$ for $0\leq
s\leq 1$, i.e., no eigenvalue that crosses the cut in the complex
plane. (It may be possible to remove this constraint, but the
Hamiltonians in the text satisfy it.)

  We first show that for infinitesimal $\Delta s$, $ \|Z(s+\Delta s)\|_2 \leq
  \|Z(s)\|_2 + \Delta s \|Y\|_2 $. We write \beq iZ(s+\Delta s)
  =\ln[e^{iZ(t)}(\one + i\Delta s Y)]\label{Z1}\enq to first
  order in $\Delta s$. Without loss of generality we can consider
  all matrices in a basis in which $Z(s)$ is diagonal, $Z(s)=$
  $\diag(\theta_1,\theta_2,\ldots,\theta_d)$, for a $d$-dimensional
  $Z(s)$, with $-\pi<\theta_j\leq \pi$. For diagonal $G$ the
  (Fr\'echet) derivative of the matrix logarithm is given by \cite
  {HornJ}\beq \frac{\partial}{\partial s}_{s \to 0}
  \ln(G+sH)=\ln^{[1]}(G)\circ H,\label{deriv}\enq where $\circ$
  signifies  the entrywise matrix product and $\ln ^{[1]}(G)$ is the
  matrix of divided differences defined as \ber (\ln^
  {[1]}(G))_{jk} & = & \frac{\ln G_{jj}-\ln
  G_{kk}}{G_{jj}-G_{kk}},~~j\neq k \label{deriv12}\\(\ln^
  {[1]}(G))_{jj} & = & 1/G_{jj}.\label{deriv11}\enr Then, putting $e^{iZ(s)}=$ $\diag
  (e^{i\theta_1},e^{i\theta_2},\ldots,e^{i\theta_d})$, for
  infinitesimal $\Delta s$, \ber iZ(s+\Delta s) & = & \ln (
  e^{iZ(s)}+ i\Delta s e^{iZ(s)}Y)\nonumber\\ & = & iZ(s)+i\Delta
  s\ln ^{[1]}(e^{iZ(s)})\circ e^{iZ(s)}Y \label{exp}\enr
  Substituting for $\ln^{[1]}(e^{iZ(s)})$ as \ber
  (\ln ^{[1]}(e^{iZ(s)}))_{jk} & = & \frac {i(\theta_j-\theta_k)}{e^{i\theta_j}-e^{i\theta_k}}
  , ~~ (j\neq k)\nonumber \\(ln^{[1]}(e^{iZ(s)}))_{jj} & = &
  e^{-i\theta_j}\label {logder2}\enr we get for the
  two-norm \ber \|Z(s+\Delta s)\|^2_2 & = & \sum _j|\theta_j+ \Delta s
  Y_{jj}|^2 \nonumber\\ & + & (\Delta s)^2\sum _{j\neq k}|\frac{i(\theta_j-\theta_k)}{e^{i\theta_j}-e^{i\theta_k}}
  e^{i\theta_j}Y_{jk}|^2.\label{Z2}\enr To first order in $\Delta
  s$ \ber \|Z(s+\Delta s)\|^2_2 & = & \sum
  _j|\theta_j|^2+2\Delta s \sum_{j}|\theta_j Y_{jj}|\label{Z3}\\ &
  \leq & \|Z(s)\|^2_2  + 2\Delta s\|Z(s)\|_2 \|Y\|_2\label{Z4}\\ & \leq &(\|Z(s)\|_2  +
  \Delta s\|Z(s)\|_2 \|Y\|_2)^2,\label{Z5}\enr
  where \er{Z4} follows from \er{Z3} by the
  Cauchy-Schwarz inequality. Thus for infinitesimal $\Delta s $ we
  have \beq \|Z(s+\Delta s)\|_2 \leq \|Z(s)\|_2 +\Delta s
  \|Y\|_2.\label{ineq2}\enq The procedure can be repeated for $\|Z(s)\|_2$ and so
forth, finally giving $\|Z(s)\|_2 \leq \|Z(0)\|_2 + s \|Y\|_2$, so
that \ber \|Z(1)\|_2 & \leq & \|Z(0)\|_2 + \|Y\|_2\label{ineq5}\\
& \equiv & \|X\|_2 + \|Y\|_2\enr which was to be proved.

(The referee pointed out that the conditions on the Hamiltonians
being traceless are not necessary.)

\end {document}